\documentclass[aps,pra,showpacs,twoside,twocolumn,10pt]{revtex4-1}
\usepackage[colorlinks=true, citecolor=blue, urlcolor=blue ]{hyperref}
\usepackage{epsfig,newlfont,amssymb,amsfonts,amsmath,bm,subfigure,palatino,mathtools,amsthm,braket,soul,enumitem,color,graphics,graphicx,times,physics}
\usepackage[normalem]{ulem}
\graphicspath{{./ab_ref/}}

\begin{document}
\title{
Designing Robust Quantum Refrigerators in Disordered Spin Models
}

\author{Tanoy Kanti Konar$^{1}$, Srijon Ghosh$^{1}$, Amit Kumar Pal$^{2}$, Aditi Sen(De)$^{1}$}
\affiliation{$^1$Harish-Chandra Research Institute,  A CI of Homi Bhabha National Institute, Chhatnag Road, Jhunsi, Prayagraj - 211019, India}
\affiliation{$^2$ Department of Physics, Indian Institute of Technology Palakkad, Palakkad 678 557, India}

\begin{abstract}

We explore a small quantum refrigerator in which the working substance is made of paradigmatic nearest-neighbor quantum spin models, the $XYZ$ and the $XY$ model with Dzyaloshinskii-Moriya interactions,   consisting of two and three spins, each of which is in contact with a bosonic bath. 
We identify a specific  range of interaction strengths which can be tuned appropriately  to 
ensure a cooling of the selected spin in terms of its local temperature in the weak- coupling limit. Moreover, we report that in this domain, when one of the interaction strengths is disordered,  the performance of the thermal machine operating as a refrigerator
remains \emph{almost} unchanged instead of degradation, thereby establishing the flexibility of this device. However, to obtain a significant amount of cooling via ordered as well as disordered spin models, we observe that one has to go beyond the weak-coupling limit and compute the figures of merits by using global master equations. 
    
\end{abstract}

\maketitle

\section{Introduction}

The quest for small quantum thermal machines~\cite{palao2001,*feldmann2003,*nimmrichter2018,*kosloff2014,*uzdin2015,*levy2012,*clivaz2019,*mitchison2019} that can supersede their classical counterparts in performance~\cite{geva1992,*feldmann2000,*ying2017,*niedenzu2018,*xu2018} has been an important and vibrant component in the field of quantum thermodynamics~\cite{gemmer2004,*kosloff2013,*kilmovski2015,*Misra2015,*millen2016,*benenti2017,*deffner2019,vinjanampathy2016,*goold2016}. These machines are expected to not only provide a better understanding of the interplay between the concepts from quantum information theory and thermodynamics~\cite{huber2015,*lostaglio2015,vinjanampathy2016,*goold2016,gour2015}, but also  lead to  build efficient quantum technologies~\cite{ikonen2017}. Moreover, the interdisciplinary nature of the designs and working principles of these machines has also attracted attention from researchers in statistical~\cite{campisi2015,*dalessio2016} and quantum many-body physics~\cite{dorner2012,*mehboudi2015,*reimann2015,*eisert2015,*gogolin2016,*skelt2019,halpern2019}.  To verify the theoretical proposals on these machines, several experiments have been performed by using trapped ions~\cite{abah2012,*rossnage2016}, mesoscopic systems~\cite{giazotto2006},  nuclear magnetic resonance~\cite{peterson2019}, and superconducting materials~\cite{karimi2016,*hardal2017,*manikandan2019}. 

Among the wide variety of small quantum thermal machines, quantum refrigerators made of quantum systems with Hilbert spaces of small dimension  have gained a lot of interest ~\cite{linden2010,skrzypczyk2011,*brunner2012,*brunner2014,*brask2015,correa2013,*correa2014,*silva2015,*naseem2020,mitchison2015,das2019,man2017,*friedman2019,*wang2015,*he2017,*du2018,*chiru2018,*seah2018,*barra2018,hewgill2020}. Special attention has recently been given to the three-spin quantum refrigerators,
where a local cooling of one of the spins is achieved by connecting each of the spins in the system with a local Markovian thermal bath. Depending on the choice of the system parameters, the refrigerator may operate in either the absorption region where energy is conserved, or in an external energy-driven region, where a channel exists between the refrigerator and an external energy source or sink.  The performance of the refrigerator and its type are assessed in terms of the heat currents between the spins and their respective baths,
 and a lowering of temperature either in the steady state or during the transient dynamics can be observed via an increase in the ground-state population of the spin undergoing local cooling~\cite{linden2010,skrzypczyk2011,*brunner2012,*brunner2014,*brask2015,correa2013,*correa2014,*silva2015,*naseem2020,mitchison2015,das2019}. Along with theoretical proposals to implement these  machines in various substrates such as quantum dots~\cite{venturelli2013}, circuit QED architectures~\cite{hofer2016}, and atom-cavity systems~\cite{mitchison2016,*mitchison2018}, three-spin quantum refrigerators have recently been implemented in laboratories using trapped ions~\cite{maslennikov2019}. 

While the original model for the three-spin refrigerator exploits a three-body interaction among the spins constituting the working substance~\cite{linden2010}, it has been shown that one can construct a three-spin refrigerator with two-body interactions also \cite{hewgill2020}, where the spin-spin interactions constitute the well-known $XXZ$ model~\cite{langari1998}, thereby highlighting the possibility of building small quantum thermal machines using paradigmatic low-dimensional quantum spin models~\cite{yang1966,orbach1958, ashida2019,zhou2017,sachdev_2011} of few spins. On one hand, it allows one to control the performance of these machines by appropriately tuning the parameters of the quantum spin Hamiltonian, which is now possible in experiments using the same substrates used for realizing thermal machines~\cite{raimond2001, duan2003,leibfried2003, cirac2004, negrevergne2006,duan2010,monz2011,pan2012}. On the other hand, existing studies on the interface of the quantum information theory and quantum spin models~\cite{fazio2008, De_Chiara2018,lewenstein2007} may prove useful in establishing the connection between quantum thermodynamics and quantum information theory. However, identifying appropriate spin Hamiltonian among numerous low-dimensional quantum spin models available in literature~\cite{zhou2017, sachdev_2011,dutta,takahasi1999} to implement a quantum refrigerator remains a demanding task.

Another challenge in implementing a working quantum refrigerator using a quantum spin model in the laboratory would be  disorder,  since imperfections are inevitably present in the system ~\cite{Shapiro2012, ahufinger2005,lee1985,abrahams1979,anderson1958}.
A disordered system  has two fundamental time-scales -- the observation time, $\tau$, over which the system undergoes a dynamics and subsequent observation via a measurement, and the time $\tau^\prime$ taken by the disordered parameter to attain its equilibrium. 
When $\tau^\prime\gg\tau$,  an effectively frozen disorder configuration during the observation time happens which can be incorporated by performing average over configurations after computing the physical quantity of interest, known as   
 quenched disordered averaging ~\cite{brout1959,de_dominicis2006,malmi2014,abaimov2015}. The realization of quantum spin models with disordered parameters being now possible in laboratories~\cite{cl2005,fort2005,fallani2007,white2009}, it is natural to ask how the performance of quantum refrigerators, built out of  quantum spin models, can alter in  presence of disorder in the system which is one of the focus of the current paper. 

In the present paper, we construct  quantum refrigerators using an one-dimensional quantum spin chain consisting of two or three  spin-$\frac{1}{2}$ particles, each of which is connected to a local Markovian bosonic thermal bath. We consider nearest-neighbour interactions among the spins, and examine a number of paradigmatic quantum spin Hamiltonian, namely quantum $XYZ$  \cite{sachdev_2011,dutta,takahasi1999} and quantum $XY$ models with  Dzyaloshinskii-Moriya (DM)  interaction \cite{moriya1960,moriya21960,anderson1959,DZYALOSHINSKY1958} as possible system Hamiltonian for the machine to operate as a refrigerator where the latter model is chosen to introduce asymmetry in the system. More specifically, we focus on two main questions as to (1) whether a small quantum refrigerator built out of quantum spin systems  always provide a significant cooling to a selected spin in terms of the population-dependent definition of local temperature, and if the answer is positive, we focus on the identification of the parameter regimes to be tuned;  and (2) whether the performance of  the  quantum thermal machine as a refrigerator remains unaffected  in the presence of quenched disorder.

We answer both questions affirmatively in terms of heat current and local temperature of the selected spins, by considering the local as well as the global master equation.
For the local master equation,  we first notice that since the magnetic fields of the initial states are aligned to the \(z\) directions, the interaction strength in the \(z\)-plane of the \(XYZ\) model have negligible effect on the refrigeration. We observe that when the couplings are weaker than the strengths of the magnetic fields, the refrigerator based on the \(XY\) model with DM interactions perform better than that of the \(XYZ\) model. Moreover,  numerical simulations reveal a small subspace of the entire parameter space in which cooling of a selected spin can take place.  Such a hierarchy remains unaltered when either the interaction strengths in the \(xy\)-plane or  the DM ones is chosen randomly from the Gaussian distribution.  Notice that although they are demonstrated by fixing the strengths of the magnetic fields,  the results remains true even for the large range of parameters.  
However, in  this domain, the refrigerator described by a quantum spin Hamiltonian,  ordered as well as disordered, does not ensure a significant cooling for a selected spin in terms of the local temperature of the spin. To overcome this, we go beyond the  local master equation and by employing global master equation, we illustrate that the local cooling provided by the ordered as well as disordered spin models, can substantially be improved. 


The rest of the paper is organized as follows. In Sec.~\ref{sec:definition}, we briefly introduce the construction of the three-spin quantum refrigerator by discussing the system Hamiltonians, the evolution of the system due to the interaction between the spins and the local Markovian bosonic baths, and the idea of local refrigeration of a selected spin during the dynamics of the system. In Sec.~\ref{sec:2spin}, we present our results on the two-spin refrigerator using ordered as well as disordered systems while we demonstrate the results for the three-spin refrigerator in Sec.~\ref{sec:three-spin}. Sec.~\ref{conclusion} bears the concluding remarks.

 \begin{figure}
     \centering
     \includegraphics[scale=0.35]{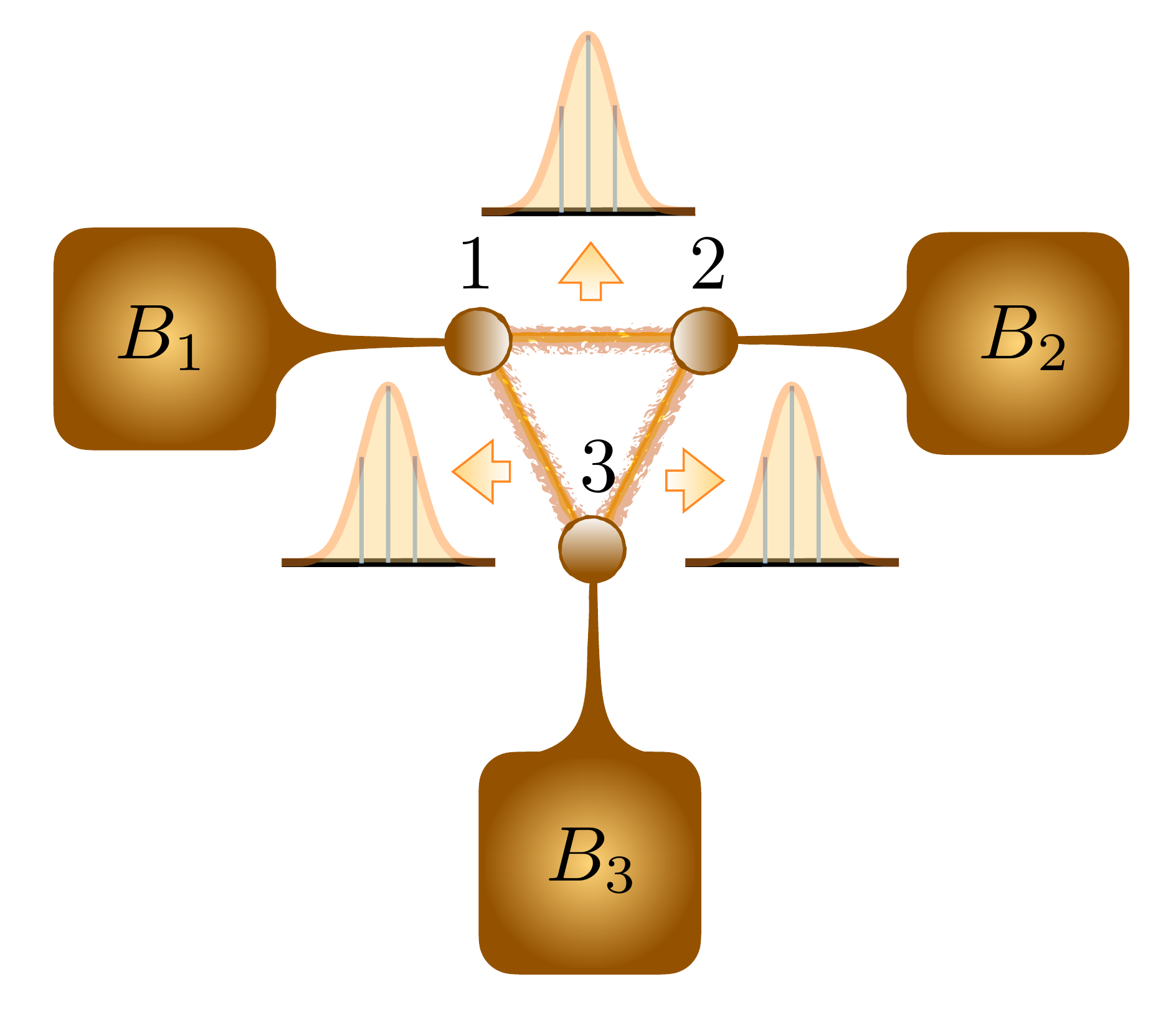}
     \caption{\textbf{A three-spin refrigerator in the presence of disorder.} Three spin-$1/2$ particles are interacting with each other via spin-exchange interactions, while individually interacting with a local thermal heat bath. The spin exchange interactions can be disordered, where the values of their strengths can be chosen from a Gaussian distributions of fixed mean and standard deviations. 
     }
     \label{fig:fig1}
 \end{figure}

\section{Quantum Refrigerator: Model and dynamics}
\label{sec:definition}

In this section, we briefly describe the quantum spin Hamiltonians used to implement a two-spin and a three-spin quantum refrigerator. The setup of the local thermal baths in contact with the individual spins, and the quantities that we have used for assessing the performance of the machine are also discussed. 

\subsection{Interacting Quantum Spin Models}
\label{subsec:iqsm}

 We model the refrigerator as an one-dimensional quantum spin chain with $N$ spin-$1/2$ particles, governed by a Hamiltonian, $H_{S} = H_{F} + H_{I}$. Here $H_{F}$ and $H_{I}=H_{xy}+H_z+H_{dm}$ correspond to the components of the system Hamiltonian $H_S$ due to the local external magnetic fields acting on each spin, and the spin-exchange interactions between the spins, respectively. They are given by 
 \begin{eqnarray}
\label{eq:hf} 
H_{F}&=&\sum_{i=1}^{N}h_{i}\sigma_{z}^{i},\\  
\label{eq:hxy}
H_{xy}&=&\sum_{i=1}^{N}J^{xy}_{i,i+1}\left[(1+\gamma)\sigma_{x}^{i}\sigma_{x}^{i+1}+(1-\gamma)\sigma_{y}^{i}\sigma_{y}^{i+1}\right],\\
\label{eq:hz}
H_z&=&\sum_{i=1}^{N}J^{z}_{i,i+1}\sigma_{z}^{i}\sigma_{z}^{i+1}, \\
\label{eq:hdm}
H_{dm}&=&\sum_{i=1}^{N}J^{dm}_{i,i+1}\left(\sigma_{x}^{i}\sigma_{y}^{i+1}-\sigma_{y}^{i}\sigma_{x}^{i+1}\right).
\end{eqnarray}
Here $\gamma$ is the $xy$ anisotropy parameter, $h_{i}$ is the strength of the local magnetic field acting on the spin $i$, $\sigma_{p}^{i}$ $(p = x,y,z)$ are Pauli matrices,  $J^{xy}_{i,i+1}$ and $J^z_{i,i+1}$ respectively represent the $xy$ and the $zz$ nearest- neighbor antiferromagnetic interaction strengths, and $J^{dm}_{i,i+1}$ denotes the strength of the Dzyaloshinskii-Moriya  interaction \cite{moriya1960,moriya21960,anderson1959,DZYALOSHINSKY1958}. Moreover, we consider  interaction strengths  to be site-independent as well as site-dependent, leading to the ordered and disordered spin systems respectively.   
A number of paradigmatic quantum spin Hamiltonian emerged from $H_S$ for different values of these system parameters are as follows.
\begin{enumerate}
\item $J_{i,i+1}^{xy},J_{i,i+1}^{dm}=0$- Classical Ising model in a parallel magnetic field, 
\item $\gamma=1$, $J^{z}_{i,i+1}=0$, $J^{dm}_{i,i+1}=0$- Transverse-field Ising model,
\item $0<\gamma<1$, $J^{z}_{i,i+1}=0$, $J^{dm}_{i,i+1}=0$- Anisotropic $XY$ model in a
transverse field, 
\item $\gamma=0$, $J^z_{i,i+1}=0$, $J^{dm}_{i,i+1}=0$- $XX$ model in a transverse magnetic field, 
\item $\gamma=0$, $J^{dm}_{i,i+1}=0$- $XXZ$ model with magnetic field, and  
\item $\gamma=0$,  $J^z_{i,i+1}=0$- $XX$ model in a transverse magnetic field with DM interaction. 
\end{enumerate}
In this paper, we focus on small quantum refrigerators, where the size is justified by the low dimension of the Hilbert space of the system. More specifically, we consider a two- and a three-spin refrigerator $(N=2,3)$ for demonstrating the results in the  subsequent sections.

\begin{figure}
    \centering
    \includegraphics[scale=0.325]{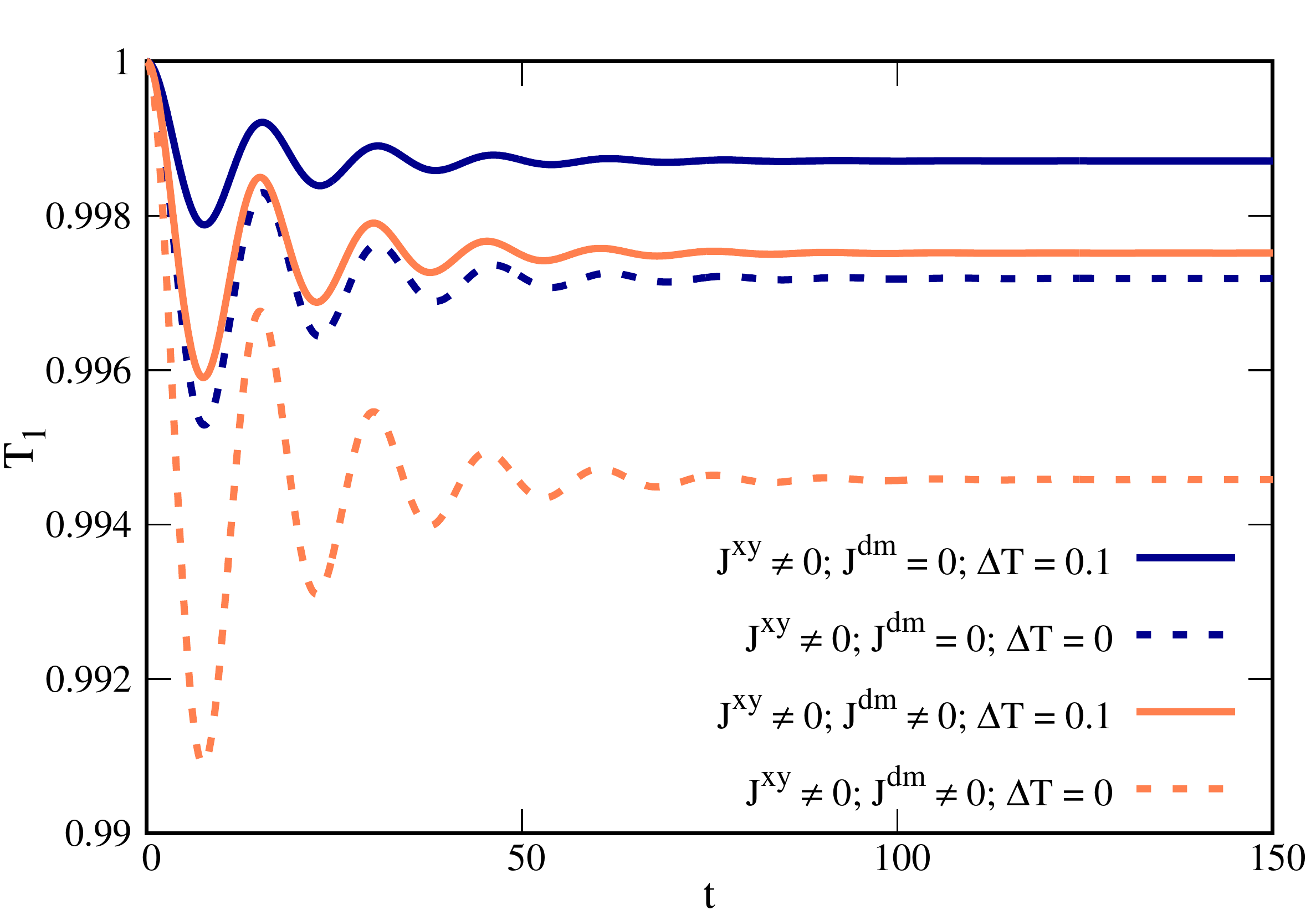}
    \caption{(Color online) \textbf{Temperature dynamics for spin $1$ of a two-spin refrigerator in weak-coupling limit.} Variation of \(T_1\) (ordinate)  vs. \(t\) (abscissa).  The initial temperatures of the two spins are $T_1(0)=1$, $T_2(0)=1.1$ (solid lines) and $T_1(0)= T_2(0)=1$ (dashed line). Dark (red) lines represent \(XX\) model with  $J^{xx}=0.02$  while light (orange) lines are for the \(XX\) model with DM interactions where $J^{xx} = J^{dm} = 0.02$. In both cases, we fix $h_1=1.1$, $h_2=1.3$, $\Gamma=0.05$, and $\gamma=0$. Both the axes are dimensionless. 
    }
    \label{fig:fig2}
\end{figure}

\subsection{Local Environments and the Open Quantum Dynamics}
\label{subsec:osd}

We now describe the system-environment setup for implementing the quantum refrigerator. We consider $N$ local heat baths, $B_1$, $B_2$, $\cdots,$ $B_N$, each of which is connected to a spin in the $N$-spin system (see Fig.~\ref{fig:fig1} for the $N=3$ case), such that any spin is completely insulated from the effect of the $N-1$ baths, except the one connected to it.
We assume that at $t=0$, the spin-exchange interactions are absent, i.e., $H_S=H_F$, and each of the spins is at thermal equilibrium with its respective environment, so that the temperature $T_i(0)$ of the spin $i$ at $t=0$ is  $T_i^0$, with  $T_i^0$ being the absolute temperature of the bath $i$. The initial state of the system, therefore, is given by $\rho_s^0=\bigotimes_{i=1}^N \rho_i^0$, where $\rho_i^0=\exp\left(-\beta_i^0 h_i\sigma_z^i\right)/\text{Tr}\left[\exp\left(-\beta_i^0 h_i\sigma_z^i\right)\right]$, with  $\beta_i^0=(k_BT_i^0)^{-1}$, $k_B$ is the Boltzmann constant. At $t>0$, all of the spin-exchange interactions, or a subset of them are turned on, so that the system is taken out of the equilibrium, and it undergoes an open system dynamics. The evolution of the state of the system, $\rho_s$, during this dynamics is described by a quantum master equation (QME) of the form  
\begin{equation}
\dot\rho_s=-\frac{\text{i}}{\hbar}[H_{S},\rho]+\mathcal{D}(\rho),
\label{eq:qme}
\end{equation} 
where $\mathcal{D}(.)$ represents the dissipator, emerging due to the spin-bath interaction. The state of the system, $\rho_s(t)$, as a function of $t$ is obtained as the solution of the QME.

\begin{figure*}
    \centering
    \includegraphics[width=0.7\textwidth]{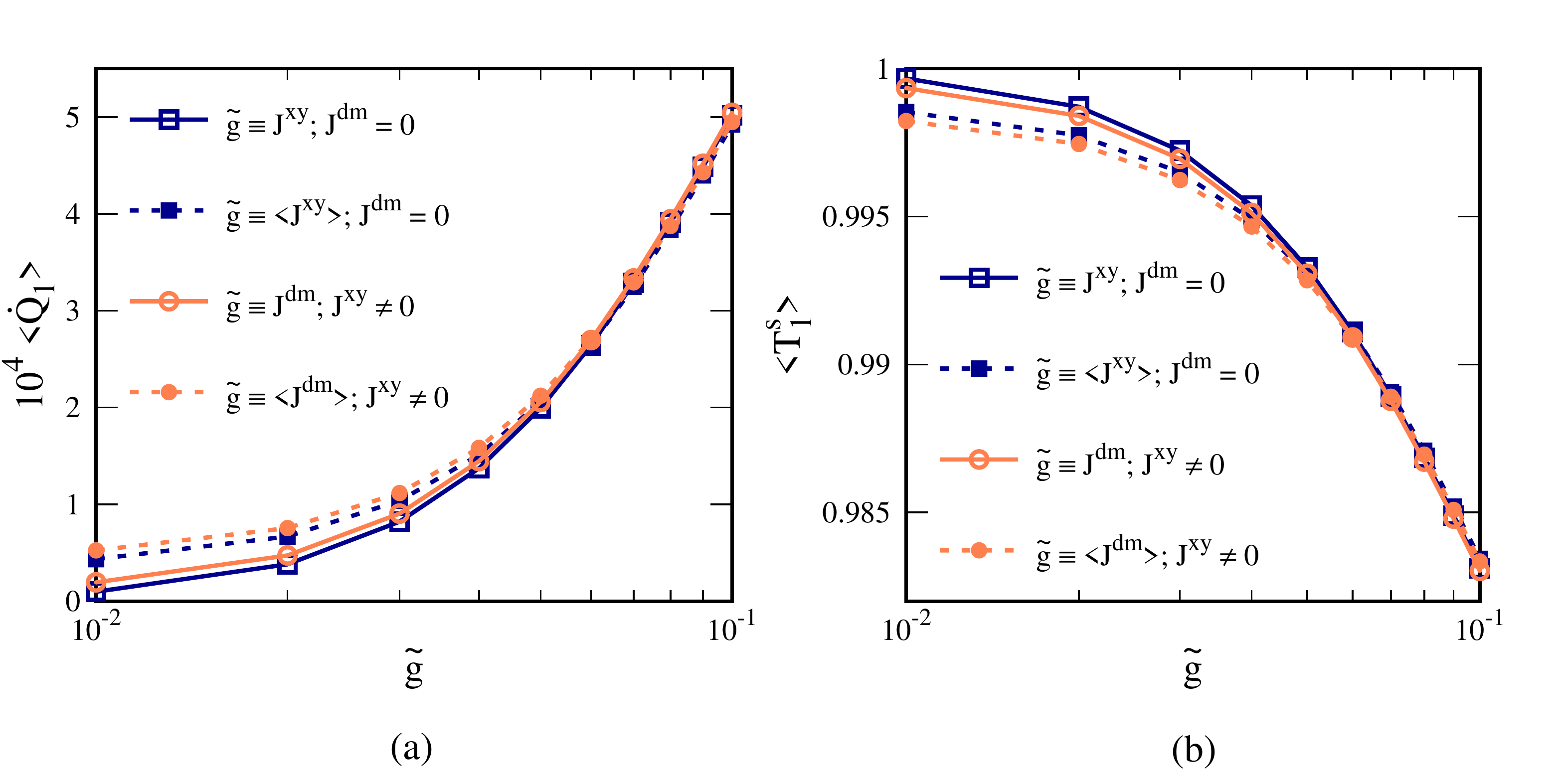}
    \caption{(Color online) \textbf{Variation of heat current and steady-state temperature (vertical axis) as functions of the strength of the spin-exchange interactions (horizontal axis).}  In figure (a) and (b),  we plot  heat current and temperature of spin $1$ with increasing  $XX$-interactions (squares) where $J^{dm}=0$ and with the increase of  DM interactions, \(J^{dm}\) (circles) having \(J^{xy} =0.02 \neq 0\).  Hollow and solid  symbols (squares as well  as circles) represent ordered and disordered spin models respectively. 
    Other parameter of the systems, namely magnetic field strengths and the spin-bath interactions are chosen as $h_1=1.1$, $h_2=1.3$ and $\Gamma=0.05$,  and the initial temperature of each spin is $T_1(0)=1$ and $T_2(0)=1.1$ respectively. Here \(\gamma =0\). All the axes are dimensionless. }
    \label{fig:fig3}
\end{figure*}

We consider each of the local thermal baths $B_i$ to be a collection of harmonic modes with a Hamiltonian $H_{b}=\int_{0}^{\omega_{m}}\dd\omega a_{\omega}^{\dagger}a_{\omega}$, where  $a_\omega$ ($a_\omega^\dagger$) is the annihilation (creation) operator corresponding to the harmonic mode of energy $\omega$, obeying $\left[a_\omega,a_{\omega'}^\dagger\right]=\delta(\omega-\omega')$, and $\omega_m$ is the maximum $\omega$.  The total interaction between the spins and their corresponding baths is represented by the Hamiltonian $H_{sb}=\sum_{i=1}^{N}\sum_\omega\left (\sigma_i^{+}\otimes a_{\omega}+\sigma_i^{-}\otimes a_{\omega}^{\dagger}\right)$, where $\sigma^{+}_{i}$ and $\sigma^{-}_{i}$ are the raising and lowering operators of the $i$-th spin respectively. The dynamical term in the QME (Eq.~(\ref{eq:qme})) takes the form~\cite{breuer2002} $\mathcal{D}(\rho)=\sum_{i=1}^N\mathcal{D}_i(\rho)$, with  
\begin{eqnarray}
\mathcal{D}_i(\rho)&=&\Gamma_i\Big[(n_\omega^i+1)\big(\sigma_i^-\rho\sigma_i^+-\frac{1}{2}\{\sigma_i^+\sigma_i^-,\rho\}\big)\nonumber\\&& +n_\omega^i\big(\sigma_i^+\rho\sigma_i^- -\frac{1}{2}\{\sigma_i^-\sigma_i^+,\rho\}\big)\Big],
\label{eq:weak_coupling}
\end{eqnarray}
in the case of the Markovian spin-bath interactions at the strict weak-coupling limit given by $h_i,\Gamma_i\gg \max\{J^{xy}_{i,i+1},J^z_{i,i+1},J^{dm}_{i,i+1}\}$. In Eq.~(\ref{eq:weak_coupling}),  $n_\omega^i$ being the occupation number of the Bose-Einstein distribution corresponding to bath $B_i$ given by $n_\omega^i=(e^{\hbar\omega/k_BT_i^0}-1)^{-1}$, with $\omega=2\hbar h_i$,  and $\Gamma_i$ being a constant. Note that the Lindblad operators represented by $\sigma^\pm_i$ here signifies local transitions among the eigenstates of the subsystem $i$, and the QME in such situations belongs to the class of local master equations. It is also important to note that in such scenarios, a violation of the second law of thermodynamics may take place, implying that a local quantum master equation may not always be appropriate to describe the stationary non-equilibrium properties of the system (see Refs.~\cite{barra2015,strasberg2017,dechiara2018}). Therefore, in the case of the local quantum master equation, the results should be interpreted carefully, and there have been proposals for rectifying this issue by constructing the master equation in a different fashion~\cite{wichterich2007}.

On the other hand, in the strong-coupling limit, the spin-interaction strengths are comparable to the strengths of the local magnetic fields, and the dynamical term corresponding to spin $i$ in  Eq.~(\ref{eq:qme}) takes the form as \cite{seah2018}
\begin{eqnarray}
\mathcal{D}_i(\rho)&=&\sum_{\omega>0}\gamma_i^\omega\Big[\big(A_\omega^i\rho A^{i \dagger}_\omega-\frac{1}{2}\{A^{i\dagger}_{\omega} A_\omega^i,\rho\}\big)\nonumber\\&& +\big(A^{i \dagger}_\omega\rho A_\omega^i-\frac{1}{2}\{A_\omega^i A^{i \dagger}_\omega,\rho\}\big)\Big],
\label{eq:strong_coupling}
\end{eqnarray}
where the operator $A_\omega^i$, given by
\begin{eqnarray}
\text{e}^{\text{i}H_S t}(\sigma^+_i+\sigma^-_i)\text{e}^{-\text{i}H_S t}=2\sum_{\omega}A_\omega^i\text{e}^{-\text{i}\omega t}
\end{eqnarray}
are the Lindblad operators on the spin $i$ corresponding to the transition of energy $\omega$ among the energy levels of the system, and is derived by decomposing the spin-part of $H_{sb}$ in the eigenbasis of $H_S$. Note that in contrast to the previous case of local master equation, the Lindblad operators here correspond to the transitions among the eigenstates of the entire system, and the QME in this situation is a global one. The coefficient $\gamma_i^\omega$ is the transition rate corresponding to the energy gap $\omega$ for the spin $i$, where  
\begin{eqnarray}
\gamma_i^\omega &=& f_i(\omega)[1+\kappa_i(\omega)],\text{ for }\omega\geq 0,\nonumber \\
\gamma_i^\omega &=& f_i(|\omega|)\kappa_i(|\omega|), \text{ for }\omega<0,
\end{eqnarray}
with $f_i(\omega)=\alpha_i\omega\text{e}^{-\frac{\omega}{\Omega}}$, with \(\Omega\) being the cut-off frequency  and $\kappa_i(\omega)=\left(\text{e}^{\hbar\beta_i\omega}-1\right)^{-1}$ representing the Ohmic spectral function and the Bose-Einstein distribution, respectively. Here, $\alpha_i$ is a constant for the  bath, $i$, quantifying the strength of the spin-bath interaction strength. In order for the Markovian approximation to be valid, we restrict the values of $\alpha_i$ such that $\max\{\alpha_i\}\ll 1$.  Here, the second law of thermodynamics is always valid. However, care must be taken while constructing quantities that are local to a subsystem of the quantum spin model. We shall elaborate on this in Sec.~\ref{subsec:three_disorder}.

\subsection{Local Refrigeration}
\label{subsec:local_cooling}

If the $N$-spin system operates as a refrigerator for the spin $i$, then the heat current, 
\begin{eqnarray}
\dot{Q}_i=\text{Tr}[H_S\mathcal{D}_i(\rho_s)],
\end{eqnarray}
corresponding to the spin $i$ in the steady state is positive~\cite{hewgill2020,seah2018,kosloff2014}. This represents a situation where heat flows from the bath $B_i$ to the spin $i$, which is at a lower temperature than $T_i^0$ in the steady state. This can also be visualized by defining a local temperature for the spin $i$~\cite{linden2010} as follows. At $t=0$, the initial state of the $i$-th spin is a diagonal state, which can be written in the eigenbasis of $\sigma_{z}$, $\{|0\rangle,|1\rangle\}$, having eigenvalues $1$ and $-1$ respectively, as $\rho_i^0=\tau_i^0\dyad{0}{0}+(1-\tau_i^0)\dyad{1}{1}$,  where $\tau_i^0=\exp(-2\beta_i^0h_i)/[1+\exp(-2\beta_i^0h_i)]$. During the dynamics, the forms of the Lindblad operators (see Sec.~\ref{subsec:osd}) ensure that the single-spin density matrix
\begin{eqnarray}
\rho_i(t)=\text{Tr}_{\underset{j,k=1,2,3}{j,k (\neq i)}}\left[\rho_s(t))\right], 
\end{eqnarray} 
at every time instant $t$, remains diagonal, i.e., $\rho_i(t)=\tau_i(t)\dyad{0}{0}+(1-\tau_i(t))\dyad{1}{1}$, while $\tau_i(t)$ varies with time starting from $\tau_i(0)=\tau_i^0$. It allows us to define a local temperature of the spin $i$ as
\begin{eqnarray}
T_i(t)=\frac{2h_i}{\ln\left[{\tau_i(t)}^{-1}-1\right]}
\label{eq:loc_temp}
\end{eqnarray}
at every time $t$, which is in agreement with the initial temperature $T_i(0)$ of the spin $i$ to be equal to $T_i^0$.

A local steady-state cooling of the spin $i$ is achieved if
\begin{equation}
T_i^s=T_i(t\rightarrow\infty)< T_i^0
\end{equation}
 at any $t>0$. Note, however, that as of now, no specific correlation between the values of $\dot{Q}_i$ and $T_i^s$ exists as we will also show here. In the subsequent sections, we demonstrate the status of the local refrigeration of a spin in the (two-) three-spin system via the heat current as well as the local temperature corresponding to the chosen spin, by appropriately tuning the system as well as the spin-bath interaction parameters.

\begin{figure*}
    \centering
    \includegraphics[width=0.7\textwidth]{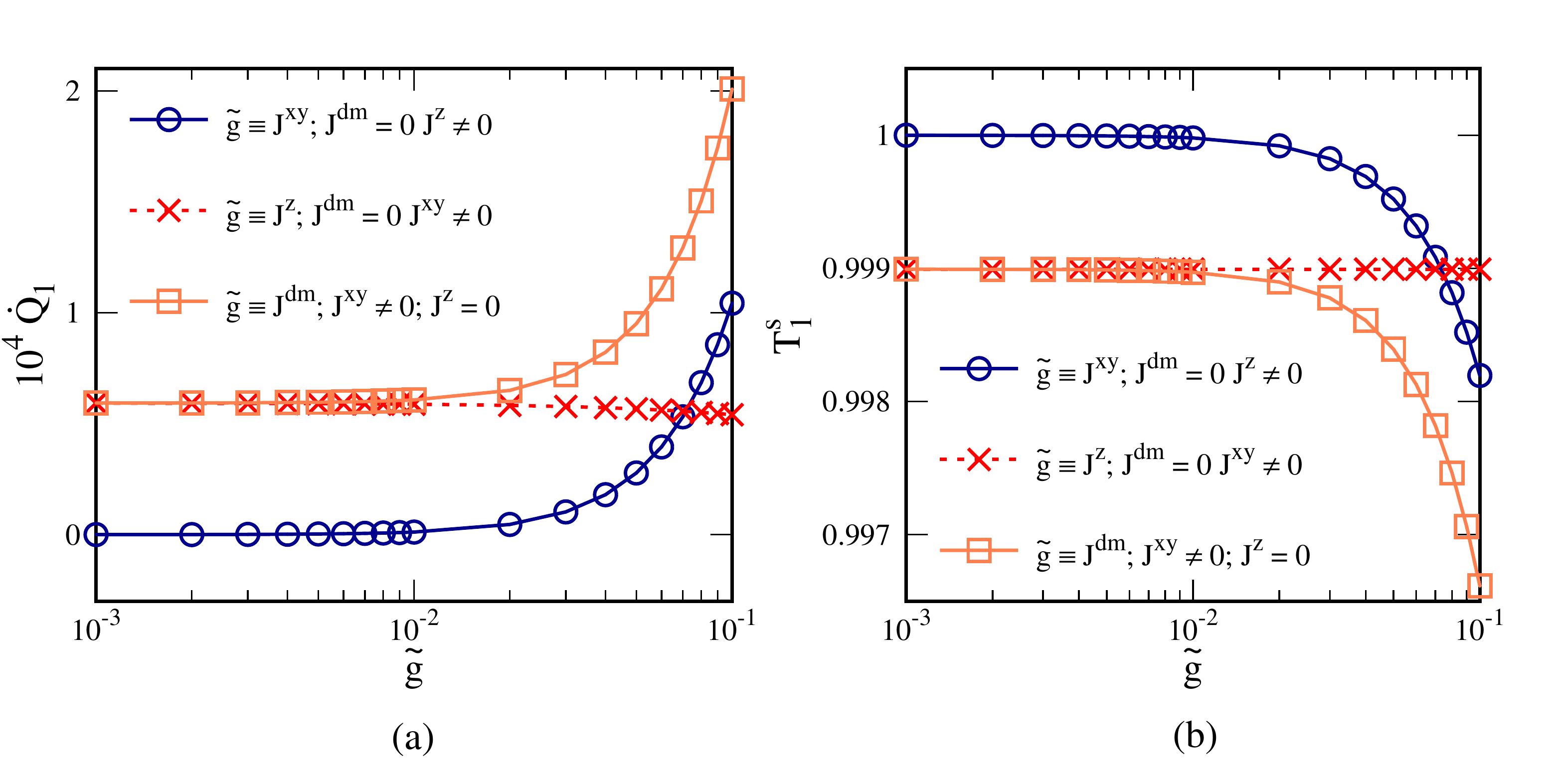}
    \caption{(Color online) \textbf{ Three-spin refrigerator: (a) $\dot{Q}_1$ and  (b) $T_1^s$ as functions of different spin-exchange interaction strengths
    where $g_{i,i+1}=g$ $\forall i\in[1,2,3]$.
    } 
    The other relevant parameters, which are kept constant, are chosen as follows. For $\tilde{g}\equiv  J^{xy}$, $J^{z}=0.019$ and $J^{dm}=0$ (circles). When $\tilde{g}\equiv J^{z}$, $J^{xy}=0.073$, and $J^{dm}=0$ (crosses) while for $\tilde{g}\equiv J^{dm}$,  $J^{xy}=0.073$,  and $J^z=0$ (squares). In all these cases, the local magnetic fields corresponding to the individual spins are fixed to $h_1=1.11$, $h_2=2.82$ and $h_3=3.65$, and the values of the spin-bath interaction parameters are $\Gamma_1=0.0639$, $\Gamma_2=0.0984$, and $\Gamma_3=0.0673$. All the axes are dimensionless.}
    \label{fig:fig4}
\end{figure*}

\section{A two-spin quantum refrigerator: Order vs. disorder}
\label{sec:2spin}

We begin our discussion with a two-spin refrigerator model (see Fig. \ref{fig:fig1} where the third spin and its corresponding bath, \(B_3\), are absent), where we focus on the local refrigeration of a chosen spin in the system. For the purpose of demonstration, we choose spin $1$ to be cooled, although the system as well as the environment parameters can be chosen appropriately to locally cool any one of the spins. To ensure that the two-spin thermal machine operates as a refrigerator for the spin $1$, we exhibit $\dot{Q}_1>0$ as well as \(T_1^s < T_1^0\) 
by properly tuning  the parameter values. Note that maintaining $\dot{Q}_1>0$ alone describes a situation that includes all the operating regimes (see Ref. \cite{hewgill2020} for three-spin refrigerator) corresponding to the two-spin thermal machine that refrigerates spin $1$.

\subsection{Ordered Spin models as Refrigerator}

\emph{Transverse $XY$ model. } Let us first consider $XY$ type spin-exchange interaction between the spins so that $H_S=H_F+H_{xy}$ for $N=2$ (see Eqs.~(\ref{eq:hf})-(\ref{eq:hxy})), where we set $\gamma=0$ for demonstration. Solving  Eq.~(\ref{eq:qme}) for the two-spin refrigerator model  via local master equation,  followed by the calculation of the local density matrix for spin $1$, leads to the local temperature of spin $1$ as
$T_1(t)=2h_1/\ln[\sigma_{11}(t)^{-1}-1]$ (see Appendix~\ref{app:appendix}). Notice that when $H_S$ represents a classical Ising model in a parallel magnetic field and the initial state of the system is a diagonal one, the system does not evolve under the  local master equation, implying that a local refrigeration of the spin $1$ is absent. Note also that under the strict weak-coupling limit (see Sec.~\ref{subsec:osd}) where the spin-interactions are negligible compared to both the local magnetic fields and the dissipation rates, our numerical analysis does not find any point in the parameter space for which a local cooling for spin $1$ can take place. This motivates us to relax the weak-coupling condition as $h_i>J^{xy}\sim\Gamma_i$ (see Ref.~\cite{mitchison2015}), where significant subspace in the parameter space of the system is found where the designed refrigerator demonstrates cooling in spin $1$. This is a feature valid for both two- and three-spin refrigerators, and from now onward, unless otherwise mentioned, we use the relaxed weak-coupling condition in terms of appropriate spin-interaction strengths (i.e, a subset of $\{J^z,J^{xy},J^{dm}\}$) to investigate the performance of refrigerators.

The observations obtained for the two-spin refrigerator modeled via a spin system other than the classical Ising model are the following: 

\begin{enumerate} 

\item A non-zero $XY$ interaction strength, \(J^{xy}\),  results in an evolution of the system, leading to a local cooling of spin $1$, irrespective of the value of $J^z$.   In Fig. \ref{fig:fig2}, the dynamics of the local temperature of the spin $1$ in a two-spin refrigerator is depicted, thereby demonstrating a local steady-state cooling. 

\item Interestingly, we find that even when $\Delta T=T_2^0-T_1^0=0$,  a steady state cooling occurs where   an energy bias is given to the system in terms of two unequal strengths of the magnetic field  to the individual spins. More importantly,  we report that vanishing $\Delta T$ proves to be advantageous with respect to cooling than that of a non-vanishing $\Delta T$ (see Ref.~\cite{ghoshal2021}) if we suitably adjust 
the parameters of \(H_s\)  and the spin-bath interaction strength (comparing solid  and  dashed lines of  Fig.~\ref{fig:fig2}). 

\item  The heat current (the steady-state temperature) remains almost constant when the strength of the spin-exchange interaction is $\leq 10^{-2}$, and increases with an increase in the value of $J^{xy}$ within the weak-coupling limit $(\leq 10^{-1})$, irrespective of the presence of the  interactions in the $z$-plane, i.e., independent of the values of \(J^z\). 
The variation of the heat current and the steady-state temperature of the spin $1$ against the strength of the spin-exchange interaction $J^{xy}$ is depicted in Figs.~\ref{fig:fig3}(a)-(b). 

\end{enumerate}

\begin{figure}
    \centering
    \includegraphics[scale = 0.6]{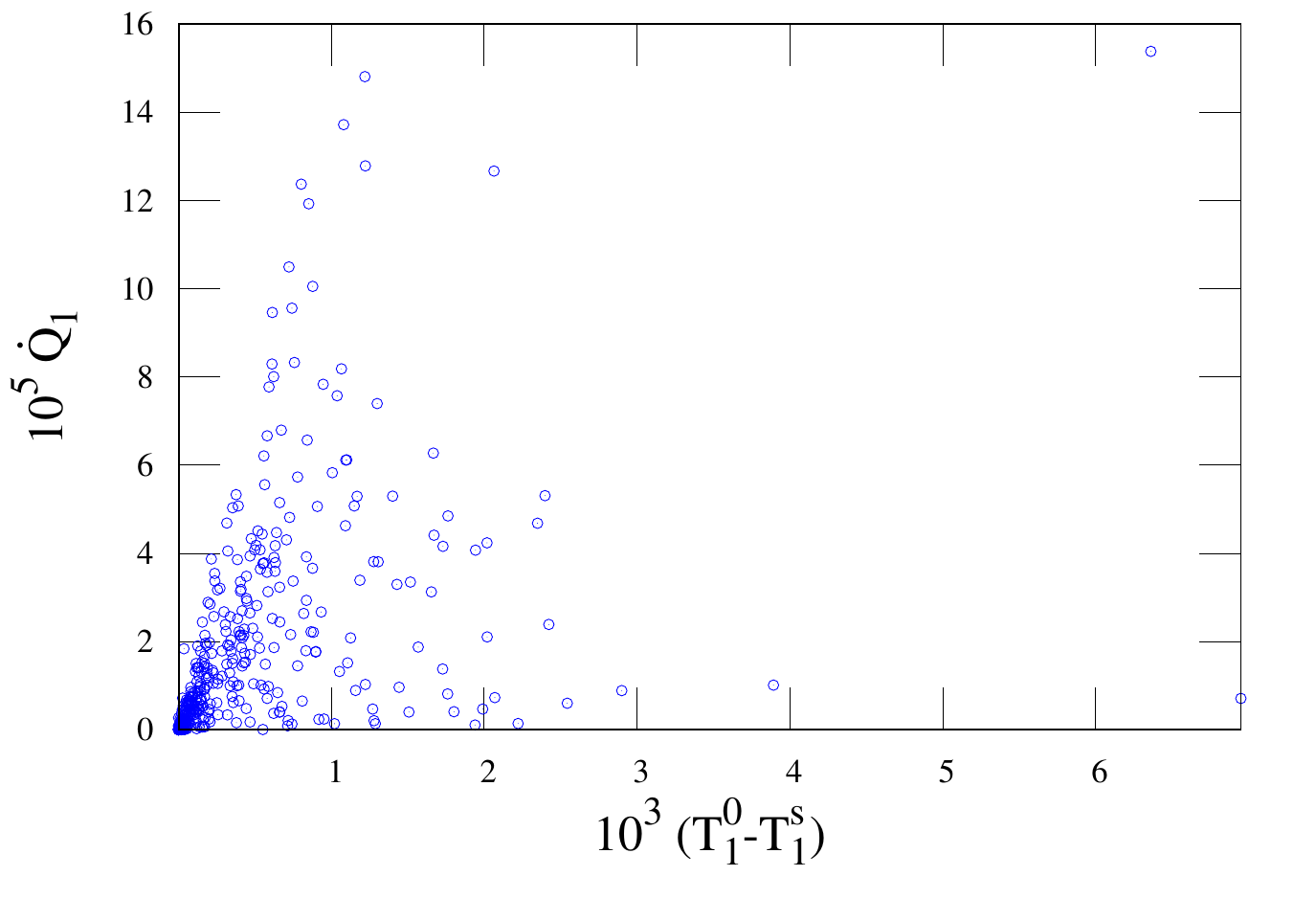}
    \caption{(Color online) \textbf{Scattered plot of $\dot{Q}_1$ (ordinate) against $T_1^0 - T_1^s$ (abscissa) of the three-spin $XXZ$ refrigerator.}  The values of the local magnetic fields, \(\{h_1, h_2, h_3\}\), corresponding to the individual spins are chosen uniformly from \([1.1,5]\)
    while the values of the spin-bath interaction parameters  $\{\Gamma_1,\Gamma_2,\Gamma_3\}$ as well as the spin-exchange interaction strengths $\{J^{xy},J^z\}$ are chosen from a uniform distribution of range $[0,10^{-1}]$. Here \(T_1^0=1\), \(T_2^0 =2\) and \(T_3^0 =3\).  Among  \(10^4\) choices of parameters, only \(4.11\%\) points are  displayed for which local temperature of the first spin is lower than unity.  
    Results indicate that there is no monotonic relation between them. Both the axes are dimensionless. }
    \label{fig:fig5}
\end{figure}

\textbf{Remark 1.} The amount of steady-state cooling achieved in the two-spin refrigerator is very small in magnitude, and it possibly  indicates that one has to go beyond  the local master equation to achieve a significant steady-state cooling of the spin $1$.

\textbf{Remark 2.} The trend remains  unchanged for $\gamma \approx 0$, with negligible effect on the amount of steady-state cooling attained during the refrigeration of spin $1$. On the other hand, when \(\gamma \rightarrow 1\), the  performance of the refrigerator diminishes. Hence the entire analysis in the rest of the paper is performed for the spin model with \(\gamma =0\).

\emph{Transverse $XY$ model with DM interaction. } To answer the question as to whether a change in the type of the spin-exchange interaction between the two spins affect the performance of the two-spin refrigerator, we add an asymmetric spin-spin interaction, specifically, the DM interaction in the system Hamiltonian, i.e., \(H_s = H_{xy} + H_{dm}\). 
We explore the behaviors of $\dot{Q}_1$ and $T_1^s$ as functions of $J^{dm}$, where $J^{xy}$ is kept fixed.

Our analysis clearly indicates that the qualitative behaviors of both the quantities, heat current as well as the steady-state temperature observed in the \emph{XX} model, remain the same even in the presence of DM interactions although  the slight improvement in terms of cooling can be seen in presence of asymmetric DM interactions, especially when  the coupling constant is weak (of the order of \(10^{-2}\)) (see Fig.~\ref{fig:fig3}) .
The local temperature dynamics of spin $1$ is shown in Fig.~\ref{fig:fig2}, while the variation of the heat current and the steady-state temperature of spin $1$ with increasing $J^{dm}$ is plotted in Fig.~\ref{fig:fig3}.

\begin{figure*}
    \centering
    \includegraphics[width=0.7\textwidth]{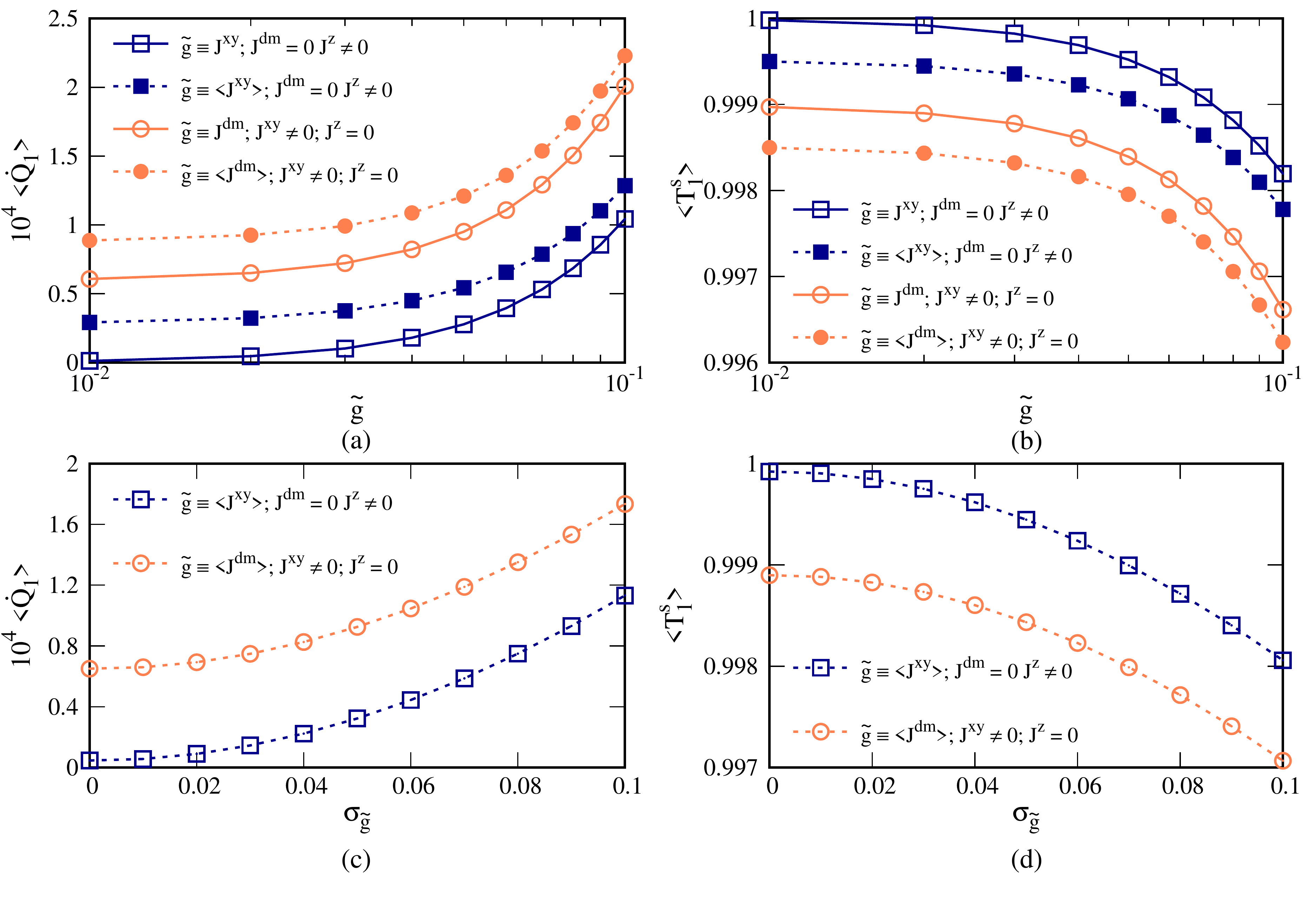}
    \caption{(Color online) 
    \textbf{ Ordered vs. disordered spin models as refrigerator} 
     (a)-(b): For $\tilde{g}\equiv \left\langle J^{xy}\right\rangle$, $J^{z}=0.019$ and $J^{dm}=0$ with $\sigma_{J^{xy}}=5\times 10^{-2}$ (dashed line with solid squares).  When $\tilde{g}\equiv \left\langle J^{dm}\right\rangle$,  $J^{xy}=0.073$, and $J^z=0$ while $\sigma_{J^{dm}}=5\times 10^{-2}$ (dashed line with solid circles). 
    The quenched averaging is performed over \(2 \times 10^3\) random configurations, chosen from Gaussian distribution with mean \(\tilde{g}\) and standard deviation, \(\sigma_{\tilde{g}}\) . 
    The similar set of parameters are also used for the ordered system (hollow circles and squares).
    All other specifications are same as in Fig. \ref{fig:fig4}.
    \textbf{(c)-(d) $\langle\dot{Q}_1\rangle$ and $\langle T_1^s\rangle$ with varying the strength of disorder, \(\sigma_{\tilde{g}}\) and $J^{dm}=0.02$.} Other specifications are similar to (a)-(b). All the axes are dimensionless. }
    \label{fig:fig6}
\end{figure*}

\subsection{Robustness in Disordered Two-spin Refrigerator}
\label{subsec:two-spin_disorder}

Let us now determine the response of the performance of the machine against  disorder in the 
two-spin refrigerator model. As mentioned in Sec. \ref{subsec:iqsm},  impurities are introduced in this model by choosing random spin-exchange interaction strengths, \(g\), from a Gaussian distribution with a  mean $\langle g\rangle$ and standard deviation $\sigma_g$, keeping the values of the local magnetic fields fixed. In this paper, either \(J^{xy}\) or \(J^{dm}\) is chosen to be 
random, by keeping 
the other coupling constants  ordered. Notice that a vanishing standard deviation reduces to  a perfectly ordered system discussed above. 

For each random parameter configuration constituted of a random value of the spin-exchange interaction strength corresponding to a random realization of the system, one can compute the quantities of interest, and subsequently take an average of the quantity over a statistically large number of parameter configurations, known as \emph{quenched} averaging of the physical quantity. Mathematically, the quenched averaging of a physical quantity, $\mathcal{Q}$, can be represented as
\begin{equation}
    \left\langle\mathcal{Q}\left(\langle g \rangle, \sigma_{g}\right)\right\rangle = \int \mathcal{P}(g)\mathcal{Q}(g)d(g),
\end{equation}
where $g$ is the parameter values of which are chosen from a Gaussian distribution ($\mathcal{P}(g)$) of mean $\langle g \rangle$ and standard deviation $\sigma_g$  quantifying the strength of the disorder.
Note that no restrictions on the possible values of the exchange interactions are imposed in order to keep the two-spin thermal machine operating in a specific working region, and a change in the values of the system parameters may, in principle, shift the two-spin thermal machine from one working region like absorption refrigerator to another such as external source driven thermal machine.

We investigate the patterns of quenched averaged heat current, $\langle \dot{Q}_1\rangle$ and steady-state temperature, $\langle T_1^s \rangle$ with  the increase of \(\langle J^{xy}\rangle \) or \(\langle J^{dm}\rangle \) where the averaging is performed over $2\times 10^3$ realizations by keeping the value of the strength of disorder fixed at $2\times 10^{-2}$.  As shown in Fig.~\ref{fig:fig3}, we  demonstrate that for small \(\langle J^{xy}\rangle \) (\(\langle J^{dm}\rangle \)), the quenched steady-state  temperature (the quenched heat current) is smaller (higher) than that obtained via ordered spin model as refrigerator. It is  also clear from the figure that the overall performance of the refrigerator remains qualitatively as well quantitatively similar in  presence of any amount of disorder in exchange interactions, thereby establishing a \emph{robustness}  of the refrigerator model against impurities.


These results provide certain insight of how a small quantum refrigerator may behave when designed using low-dimensional quantum spin Hamiltonian, and when disorder is present in the system. However, it is not clear whether these trends remain the same when one considers the  traditional three-spin refrigerator. We explore this in the next section.

\section{Three-Spin Refrigerator based on Quantum Spin Model}
\label{sec:three-spin}

In order to check whether the results of the two-spin refrigerator remains qualitatively valid also for the widely studied three-spin refrigerator, we first explore the case of identical spin-exchange interactions between all spins, i.e, $g_{i,i+1}=g$ $\forall i\in[1,2,3]$, where $g$ stands for  different types of spin-exchange interactions (see Secs.~\ref{subsec:iqsm} and ~\ref{subsec:two-spin_disorder}). For brevity, we denote $J^{xy}_{i,i+1}=J^{xy}$, $J^{z}_{i,i+1}=J^{z}$, and $J^{dm}_{i,i+1}=J^{dm}$ for all $i$.

Unless otherwise stated, we assume the constraint $T_1^0\leq T_2^0\leq T_3^0 $ for the bath temperatures, and always choose their values as $T_1^0=1$, $T_2^0=2$, $T_3^0=3$ for demonstration. By fixing the strengths of the magnetic fields, we study the response of the machine on the local cooling phenomena, specifically  in terms of  $\dot{Q}_1$ as well as \(T_1^s\), when interaction strengths, \(J^{xy}\), \(J^z\), and \(J^{dm}\) are varied in the range $[10^{-3},10^{-1}]$ (see  Fig.~\ref{fig:fig4}).  Notice that a stark difference between the two- and the three-spin refrigerators is that for the latter, there are possibilities to choose different interaction strengths between spins, \(i\) and \(i+1\), \(i =1, 2, 3\). In this work, we take them to be site-independent although site-dependence does not substantially effect the cooling procedure  as we will see in the succeeding subsection.  

\emph{Role of interaction strength on refrigeration.} The observations for the three-spin refrigerators are quite similar to the two-spin ones and can be divided into three categories -- (1)  increase of \(J_z\) while \(J^{xy} \ne 0\), \(J^{dm} =0\); (2) variations of \(J^{xy}\) with fixed \(J^z\) and \(J^{dm} =0\), leading to the \(XYZ\)-refrigerator; (3) change of \(J^{dm}\) by fixing \(J^{xy}\) with \(J^z=0\) which can be referred as the \(XYDM\)-refrigerator.   
In the first case, the presence of a non-zero $xy$ interaction in the system results in a slow variation of $\dot{Q}_1$ with $J^z$, while the corresponding change in the steady state temperature $T^s_1$ of spin $1$ is vanishing (see Fig.~\ref{fig:fig4}(b) for the behavior of $T^s_1$ corresponding to the data presented in Fig.~\ref{fig:fig4}(a)).  The increase (decrease) of $\dot{Q}_1$  ($T_1^s$) becomes more prominent in the second and the third scenarios. As pointed out in the case of two-spin refrigerator, the  refrigeration  can be improved by  varying DM interaction strength compared to the \(XXZ\)-refrigerator as depicted in Fig. \ref{fig:fig4}. 
In all these calculations, we fix $\gamma=0$ in $H_{xy}$ (i.e., $XX$ model) since our data suggests that a non-zero value of $\gamma$ in the neighborhood of \(XX\) model has no significant effect on the refrigeration of spin $1$ and the performance of the refrigerator degrades with the increase of \(\gamma\). 

\begin{figure*}
    \centering
    \includegraphics[width=0.7\textwidth]{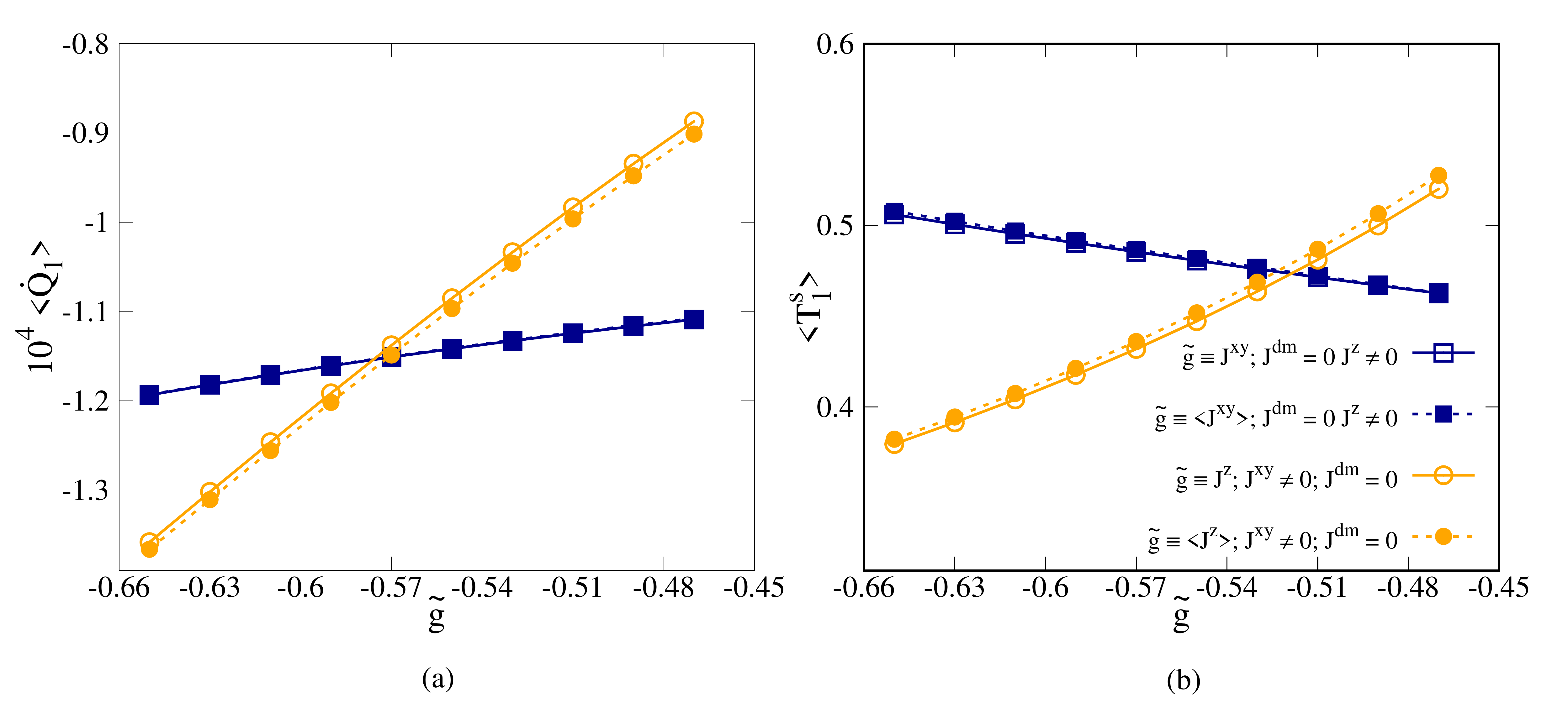}
    \caption{(Color online) \textbf{Study of refrigeration with global master equation.} (a) \(\dot{Q}_1\) for the ordered spin model and  $\langle\dot{Q}_1\rangle$ for the disordered ones vs. \(\tilde{g}\). (b) 
    Steady-state temperature and its quenched averaged one with varying interaction strengths. 
    Both for disordered and ordered situations, when $\tilde{g}\equiv J^{xy}$ or \(\equiv \langle J^{xy}\rangle\), $J^{z}=-0.55$ and $J^{dm}=0$ (solid squares for disordered and hollow squares for ordered) while $\tilde{g}\equiv J^{z}$, or \(\equiv \langle J^{z}\rangle\),  $J^{xy}=-0.4$, and $J^{dm}=0$ (solid circles and hollow circles for disordered and ordered respectively). Initial temperatures are same as in other three-spin refrigerators.
    Here
    $h_1=0.1$, $h_2=1.5$, $h_3=1.4$, and  $\alpha_1=10^{-4}$, $\alpha_2=10^{-3}$, and $\alpha_3=10^{-2}$. In the disordered-case, averaging is performed over \(5 \times 10^2\) configurations. All the axes are dimensionless.  }
    \label{fig:fig7}
\end{figure*}

As it is clear from  Figs.~\ref{fig:fig4}(a) and (b),  there is little or no variation of $\dot{Q}_1$ and $T_1^s$ as a function of the spin exchange interactions, when the interaction strength is $\leq 10^{-2}$. Beyond $10^{-2}$, the variations of $\dot{Q}_1$ and $T_1^s$ increase with increasing the spin-exchange interaction strength.  Also, it is important to note that in the strictly weak-coupling regime, the local refrigeration obtained in spin $1$ is negligible, although the three-spin machine operates in the refrigerator region for spin $1$. These findings suggest that in order to obtain a significant cooling in terms of the temperature of spin $1$, one needs to explore beyond the local master equation, as was also indicated by the results on the two-spin refrigerator. To investigate whether significant cooling can be found beyond this local master equation domain, we relax the weak-coupling condition to $h_i>\max\{J^{xy},J^z,J^{dm}\}$, and find that a considerable steady state cooling may indeed be present in such situations. See Fig.~\ref{fig:fig7} for a typical example, where we have set $J^{xy},J^{z} \ne 0$ and $J^{dm}=0$.

\emph{Connecting heat current with local temperature in three-spin model based refrigerator.} Let us here address the question -- \emph{whether a high positive value of $\dot{Q}_1$ always implies a low value of steady-state temperature in a specific spin.}. To demonstrate it, 
we choose $10^4$ random parameter configurations of the three-spin refrigerator, where the system Hamiltonian is represented by $H_S=H_F+H_{xy}+H_{z}$, and we assume $g_{i,i+1}=g$ $\forall i\in[1,2,3]$, where $g\equiv J^{xy},J^z$. The random values of the spin-exchange interaction strengths,
and the spin-bath coupling strengths $\Gamma_i$, $\forall i\in[1,2,3]$, are  chosen from a uniform distribution within $[0,10^{-1}]$. In the scatter diagram presented in Fig.~\ref{fig:fig5}, each point represents a three-spin thermal machine performing local refrigeration for spin $1$, which is indicated by $T_1^0-T_1^s>0$ and $\dot{Q}_1>0$. It is clear from the corresponding amounts of the steady-state cooling that no specific correlation exists between $T_1^0-T_1^s$ and $\dot{Q}_1$. Specifically, a very low value  of heat current can lead to a substantially low  steady-state temperature and vice-versa.   Note also that only about $4.11\%$ of the $10^4$ randomly chosen points result in $\dot{Q}_1>0$, which remains almost unchanged even in the presence of an additional DM term in $H_S$ (in this case, the percentage is $3.25\%$). It  again indicates the scarcity of a working three-spin refrigerator providing a significant amount of cooling by considering local master equation, which indicates the importance of  identifying the subspace in the entire parameter space for designing a small quantum refrigerator using the chosen quantum spin models.

\subsection{Disorder-enhanced Refrigeration in  Three-spin Systems}
\label{subsec:three_disorder}

We will now examine how impurities arising naturally in the spin model affect the refrigeration. To incorporate impurities in this three-spin refrigerator model, interaction strengths, i.e.,  \(J^{xy}_{i, i+1}\) and \(J^{dm}_{i, i+1}\) are taken to be site-dependent and are chosen randomly from the Gaussian distribution with mean, \(\langle J^{xy}\rangle\), and \(\langle J^{dm}\rangle\)) having   standard deviation, \(\sigma_{J^{xy}}\) and \(\sigma_{J^{dm}}\) respectively. The magnetic fields are fixed to the same value mentioned in the ordered case (see Fig. \ref{fig:fig4}).  Finally we compute the quenched averaged heat current,   $\langle \dot{Q}_1\rangle$ and quenched steady-state temperature, $\left\langle T_1^s\right\rangle$ of spin $1$  by averaging over $2 \times 10^3$ random configurations for a given  strength of the disorder. Both with the random $XY$ as well as DM interaction strength, i.e., for a given \(\langle J^{xy} \rangle\) or \(\langle J^{dm} \rangle\) and their corresponding \(\sigma_{J^{xy}}\) (\(\sigma_{J^{dm}}\)), we report that  
\begin{equation}
\langle \dot{Q}_1\rangle > \dot{Q}_1 \, \, \mbox{and} \left\langle T_1^s\right\rangle <   T_1^s, 
\end{equation}
which establishes the \emph{disorder-induced  thermal device} although the increase (decrease) of heat current (temperature of the first spin) is small.  It should be noted that  although in Figs.~\ref{fig:fig6}(a)-(b), we depict the enhancement of cooling feature by using disordered three-spin refrigerator over its ordered counterparts by choosing exemplary values of magnetic fields and other interaction strengths, the characteristics remain same even for other range of parameters in the local master equation. Therefore, as argued in case of two-spin refrigerator, our analysis clearly indicates that the spin model as thermal machine is  robust against impurities. 

A comment on the significance of the enhancement of the cooling phenomena in the disordered refrigerator is in order here. For brevity of the notation, let us again denote the disordered spin-interaction strength by $g$, where in the present work, we choose $g$ to be either $J^z$, or $J^{xy}$ (see also Sec.~\ref{subsec:two-spin_disorder}, and Figs.~\ref{fig:fig3},\ref{fig:fig4},\ref{fig:fig6}, and \ref{fig:fig7}). Let us denote by $g_0$ the value of $g$ for which 
\begin{eqnarray}
\dot{Q}_1(g_0)&=&\max\dot{Q}(g),\nonumber \\  
T^s_1(g_0)&=&\min T_1^s(g),\nonumber  
\end{eqnarray}
where the maximization and minimization is performed over the entire range of $g$ satisfying the weak-coupling constraint, and by definition, $\langle\dot{Q}_1\rangle\leq \dot{Q}_1(g_0)$ and $\langle T^s_1\rangle\geq T^s_1(g_0)$. This can interpret the results reported in Figs.~\ref{fig:fig6}(c)-(d) as being far from the optimal value $g_0$ of $g$. Note, however, that under local master equation, $\dot{Q}_1$ ($T^s_1$) increases (decreases) monotonically with $g$, and $g_0$ is the point $g_0=10^{-1}$ in the chosen range of $g$. While finding $\langle\dot{Q}_1\rangle \leq \dot{Q}_1(\langle g\rangle)$ ($\langle T^s_1\rangle \geq T^s_1(\langle g\rangle)$) is likely for such monotonically increasing (decreasing) behaviour of $\dot{Q}_1$ ($T^s_1$) when  $\langle g\rangle$ is far from $g_0$, such straightforward predictions can not be made for quantities that vary non-monotonically with $g$. This highlights the importance of investigating the possibility of enhancement (decrease) in the value of $\dot{Q}_1$ ($T^s_1$).

\emph{Effects of strength of disorder on refrigeration.} To probe further, let us check the role of the magnitude of the disorder on the observed robustness. We systematically increase the value of the disorder-strength up to $10^{-1}$, and observe that with increasing strength of the disorder, the average value of the heat current of the first spin attains a more positive value, while the steady-state temperature becomes lower (see Figs.~\ref{fig:fig6}(c)-(d)) than that of the model with low disorder-strength. It clearly exhibits an advantage to attain a lower steady-state temperature of the refrigerated spin in the presence of disorder where one is forced to operate a small quantum thermal machine made of three spins as a refrigerator.

\emph{Beyond weak-coupling limit.}
All the results obtained till now strongly pinpoints that  spin-exchange interaction strength beyond the weak-coupling limit aids in attaining a lower steady-state temperature of the refrigerated spin. This poses the natural question as to whether a quantum refrigerator in the strong-coupling domain performs advantageously to obtain an even lower steady-state temperature. It is also logical to ask whether the robustness of the three-spin refrigerator against  disorder remains unaltered in the strong coupling regime. Our numerical study of the three-spin refrigerator in the strong coupling limit using the global master equation, as described in Sec.~\ref{subsec:osd}, answers both the questions positively. 

Both in ordered as well as disordered scenarios, we find that the  steady-state temperature and  the corresponding  quenched averaged temperature of the first spin can  substantially be decreased  in the strong-coupling domain compared to that obtained in the weak-coupling limit.  
In Fig.~\ref{fig:fig7},  the patterns of the steady state temperature, $T_1^s$ as well as \(\langle T_1^s \rangle\) by varying the corresponding interaction strengths, \(J^{xy}\) or \(J^z\), are depicted by fixing local magnetic fields of all the spins comparable to the coupling constants. Note here that due to numerical limitations, we perform here quenched averaging over \(5\times 10^2\) configurations.  
In this regime also, we exhibit that effects of randomness in interaction strengths on the physical quantities quantifying the performance of the thermal machine is not significant, thereby supporting our claim of the robustness of the  quantum refrigerator against quenched disorder. 

While the robustness of local cooling in the disordered refrigerator is a common feature in both 
 local and global master equations, an interesting difference between these two situations emerge from Fig.~\ref{fig:fig7}. Note that  in the ordered case, a lower steady-state temperature for spin $1$ can be obtained by varying $J^z$ for a fixed value of $J^{xy}$, compared to the situation when $J^{xy}$ is varied keeping $J^z$ fixed. The situation alters after a certain threshold value of the varying parameter.
 A higher enhancement of cooling, in terms of both heat current as well as local temperature of spin $1$, is also obtained when disorder is present in $J^z$, compared to when $J^{xy}$ is disordered. These observations indicate that $J^z$ occasionally outperforms $J^{xy}$ in enhancing the performance of the refrigerator. In the same context, note that the results reported on the weak-coupling range of the spin-interaction strengths remain invariant under changing the value of $J^z$ from a zero to a non-zero value. However, under  global master equation, the performance of the refrigerator depends qualitatively (i.e., in terms of presence or absence of cooling) as well as quantitatively (i.e., in terms of the amount of cooling obtained) on the value of $J^z$. This is justified by the result that for a fixed non-zero value of $J^{xy}$ (for instance, when $-0.65\leq J^{xy}\leq -0.45$), the system may also exhibit a steady-state heating of spin $1$ at $J^{z}=0$, and a local cooling of spin $1$ starts to appear only when $J^z\leq J^z_c$, where $J^z_c$ is a critical value of $J^z$ that depends on the chosen value of $J^{xy}$.

Before concluding, let us  point out that the heat current for spin $1$ in the strong-coupling scenario is negative, which is in contrast to a positive heat current expected for a spin, undergoing a local cooling. Note that the strong-coupling scenario corresponds to a global approach of constructing the quantum master equation (see Sec.~\ref{subsec:osd}). In view of this, one needs to be careful in defining the heat current, since a definition in terms of the local Hamiltonian, given by $\dot{Q}_i=\text{Tr}(H_F^i \mathcal{L}_i(\rho))$, where $H_F^i$ and $\mathcal{L}_i(\rho)$ are respectively the local Hamiltonian and the dissipating term corresponding to the subsystem $i$, may not be appropriate for the validity of the balance equation given by
\begin{equation}
\label{balance}
\Delta=\frac{dS}{dt}-\sum_i \frac{Q_i}{k_BT_i},
\end{equation}
which, in turn, ensures the validity of the second law of thermodynamics~\cite{barra2015,strasberg2017,dechiara2018,ghoshal2021}. Here, $\Delta$ and $S$ respectively are the entropy production rate and the entropy of the system, $Q_i$ is the heat-flow from the system to the $i$th bath, $k_B$ is the Boltzmann constant, and $T_i$ is the absolute temperature of the bath $i$. This implies that the determination of $\dot{Q}_i$ requires a careful analysis (see, for example, Ref.~\cite{hewgill2021}), and in an effort to avoid the inconsistency arising from defining the heat currents using the local Hamiltonian, we have used the full system Hamiltonian $H_S$, including both the local and the interaction parts, to define the heat current as $\dot{Q}_i=\text{Tr}(H_S \mathcal{L}_i(\rho))$.  It is important to stress here that although one is interested in the local properties of the refrigerator, in a global approach, the dynamics of the system is determined as a whole, and extracting information about a specific subsystem is non-trivial due to the strong interactions between individual subsystems. However, this does not affect the main thesis of this paper, since local cooling of spin $1$ is seen in both cases of the local and global master equation approach.


\section{Conclusion}
\label{conclusion}

A potential method to build  a small scale quantum  thermal machines is via quantum spin models which can be implemented by using physical substrates like  trapped ions and neutral atoms in optical lattices. We chose this avenue to design  quantum refrigerators consisting of two and three spins based on nearest-neighbour quantum \(XYZ\) model as well as quantum \(XY\) model with  DM interactions. The initial state of the device is prepared in the thermal equilibrium states of the individual spins  which are attached with their respective local baths, and their interactions are turned on during the dynamics which  is the refrigeration process. In this paper,  the interaction strength is considered to be both ordered as well as disordered.   Our aim  is to show the  reduction of local temperature in one of spins at the steady state, thereby exhibiting the refrigeration. We call this device to be a refrigerator when the temperature of that spin is lower than the minimum of the initial temperatures of all the spins.

By considering the local master equation, we found that the cooling of one of the spins occurs when the parameters of the ordered spin models are appropriately tuned. Specifically, we observed that DM interactions help to reach lower temperature than that of the \(XYZ\) model while interactions in the \(z\)-plane of the \(XYZ\) model does not help at all. 
During the preparation procedure of the spin model, it is quite natural to have impurities in the system and hence refrigeration should  be effected by the disorder. We observed that both in two- and three-spin refrigerator models, instead of decreasing the performance, disorder in the interaction strength can help to increase the figures of merits for refrigeration, although the advantage is not significant. It clearly illustrates that the spin model-based quantum thermal machines are robust against impurities. We finally showed that the robustness against disorder can also be confirmed beyond the weak-coupling limit which is by investigating the global master equation. 
In future, it will be interesting to study  whether the robustness observed against disorder on quantum spin model-based thermal devices remains valid for other spin models having different intricacies.

\acknowledgements

TKK, SG, and ASD acknowledge the support from the Interdisciplinary Cyber Physical Systems (ICPS) program of the Department of Science and Technology (DST), India, Grant No.: DST/ICPS/QuST/Theme- 1/2019/23. AKP acknowledges the Seed Grant from IIT Palakkad. We  acknowledge the use of \href{https://github.com/titaschanda/QIClib}{QIClib} -- a modern C++ library for general purpose quantum information processing and quantum computing (\url{https://titaschanda.github.io/QIClib}), and the cluster computing facility at the Harish-Chandra Research Institute. We also thank the anonymous Referee for valuable suggestions.

\appendix

\section{Quantum master equation for the two-spin model}
\label{app:appendix}

For a two-spin model, let us consider the general form of the density matrix at time $t$, given by 
\begin{equation}
\rho(t) =  \begin{bmatrix}
\rho_{11}(t) & \rho_{12}(t) & \rho_{13}(t) & \rho_{14}(t)\\
\rho_{21}(t) &\rho_{22}(t) & \rho_{23}(t) & \rho_{24}(t)\\
\rho_{31}(t) & \rho_{32}(t) & \rho_{33}(t) & \rho_{34}(t)\\
\rho_{41}(t) & \rho_{42}(t) & \rho_{43}(t)) & \rho_{44}(t)
\end{bmatrix},  
\end{equation}
where $\rho_{ij}(t)=a_{ij}(t)+\text{i} b_{ij}(t),$ $\forall$ $i \ne j$ and $\rho_{ii}(t)=a_{ii}(t),$ $\forall$  $i=j$, both $a_{ij}(t)$ and $b_{ij}(t)$ being real. Consider the initial state of the system to be $\rho^0=\rho_1^0\otimes \rho_2^0$, where $\rho_i^0=\tau_i^0\dyad{0}{0}+(1-\tau_i^0)\dyad{1}{1}$  with $\tau_i^0=\exp(-2\beta_i^0h_i)/[1+\exp(-2\beta_i^0h_i)]$, $i=1,2$. Time-evolution of this state, according to Eqs.~(\ref{eq:qme})-(\ref{eq:weak_coupling}), with \(H_S = H_F + H_{xy}\) (\(\gamma =0\)), can be determined by solving the 16 coupled differential equations, given by 
\small
\begin{widetext}
\begin{eqnarray*}
\dot{a}_{11} &=& \Gamma [a_{33} n^{1}_{2h_1} - a_{11}(2+n^{1}_{2h_1}+n^{2}_{2h_2})+a_{22} n^{2}_{2h_2}];\;\; \dot{a}_{12} = \Gamma[-a_{12}(1.5 +n^{1}_{2h_1}+n^{2}_{2h_2}) + a_{34} n^{1}_{2h_1}] - 2 b_{13}J + 2 b_{12} h_2;\\
\dot{b}_{12} &=& \Gamma[-b_{12}(1.5 +n^{1}_{2h_1}+n^{2}_{2h_2}) + b_{34} n^{1}_{2h_1}] + 2 a_{13}J - 2 a_{12}h_2;\;\;
\dot{a}_{13} = \Gamma[-a_{13}(1.5+n^{1}_{2h_1}+n^{2}_{2h_2})+a_{24}n^{2}_{2h_2}]-2b_{12}J+2b_{13}h_1;\\
\dot{b}_{13} &=& \Gamma[-b_{13}(1.5+n^{1}_{2h_1}+n^{2}_{2h_2})+b_{24}n^{2}_{2h_2}]+2a_{12}J-2a_{13}h_1;\;\;
\dot{a}_{14} =-\Gamma a_{14}(1+n^{1}_{2h_1}+n^{2}_{2h_2})+2b_{14}(h_1+h_2);\\
\dot{b}_{14} &=& -\Gamma b_{14}(1+n^{1}_{2h_1}+n^{2}_{2h_2})+2a_{14}(h_1+h_2);\;\;
\dot{a}_{22} = \Gamma[a_{11}(1+n^{2}_{2h_2})-a_{22}(1+n^{1}_{2h_1}-n^{2}_{2h_2})+a_{44}n^{1}_{2h_1}]-4b_{23}J;\\
\dot{a}_{23} &=& -\Gamma a_{23}(1+n^{1}_{2h_1}+n^{2}_{2h_2})+2b_{23}(h_1-h_2);\;\;
\dot{b}_{23} =-\Gamma b_{23}(1+n^{1}_{2h_1}+n^{2}_{2h_2})+2J(a_{22}-a_{33})-2a_{23}(h_1-h_2);\\
\dot{a}_{24}&=&\Gamma[a_{13}(1+n^{2}_{2h_2})-a_{24}(0.5+n^{1}_{2h_1}+n^{2}_{2h_2})]+2b_{34}J+2b_{24}h_1;\;\;
\dot{b}_{24} =\Gamma[b_{13}(1+n^{2}_{2h_2})-b_{24}(0.5+n^{1}_{2h_1}+n^{2}_{2h_2})]-2a_{34}J-2a_{24}h_1;\\
\dot{a}_{33}&=&\Gamma[a_{11}(1+n^{1}_{2h_1})-a_{33}(1+n^{1}_{2h_1}+n^{2}_{2h_2})a_{44}n^{2}_{2h_2}]+4b_{23}J;\;\;
\dot{a}_{34}=\Gamma[a_{12}(1+n^{1}_{2h_1})-a_{34}(0.5+n^{1}_{2h_1}+n^{2}_{2h_2})]+2b_{24}J+2b_{34}h_2;\\
\dot{b}_{34}&=&\Gamma[b_{12}(1+n^{1}_{2h_1})-b_{34}(0.5+n^{1}_{2h_1}+n^{2}_{2h_2})]-2a_{24}J-2a_{34}h_2;\;\;
\dot{a}_{44} = \Gamma[a_{22}(1+n^{1}_{2h_1})+a_{33}(1+n^{2}_{2h_2})-a_{44}(n^{1}_{2h_1}+n^{2}_{2h_2})].
\end{eqnarray*}
\end{widetext}\normalsize 
with $n^{1}_{2h_1}=1/(\exp(2\beta_1^0 h_1)-1)$ and  $n^{2}_{2h_2}=1/(\exp(2\beta_2^0 h_2)-1)$ (see Eq.~(\ref{eq:weak_coupling}) and the following discussion.). Notice that the above coupled differential equations will be changed when \(H_s = H_F + H_{xy} + H_{dm}\). The  time-dependent density matrix $\rho(t)$ of the two-spin system reads as
\begin{equation}
\rho_{s}(t) = \begin{bmatrix}
\rho_{11}(t) & 0 & 0 & 0\\
0 &\rho_{22}(t) & \rho_{23}(t) & 0\\
0 & \rho_{32}(t) & \rho_{33}(t) & 0\\
0 & 0 & 0 & \rho_{44}(t)
\end{bmatrix}.
\end{equation}
Tracing out spin $2$, the local density matrix of spin $1$ takes the form,
\begin{equation}
\rho_{1}(t) = \begin{bmatrix}
\sigma_{11}(t) & 0\\
0 &\sigma_{22}(t)\\
\end{bmatrix}, 
\end{equation}
where $\sigma_{11}(t)=\rho_{11}(t)+\rho_{22}(t)$ and $\sigma_{22}(t)=\rho_{33}(t)+\rho_{44}(t)$.

\bibliography{ref.bib}

\end{document}